\documentclass[manuscript]{aastex}

\begin{document}

\title{Study the build-up, initiation and acceleration of 2008 April 26 coronal mass ejection observed by STEREO}

\author{X. Cheng\altaffilmark{1,2}, M. D. Ding\altaffilmark{1,2} and J. Zhang\altaffilmark{3,1}}

\altaffiltext{1}{Department of Astronomy, Nanjing University,
Nanjing, 210093, China; dmd@nju.edu.cn}

\altaffiltext{2}{Key Laboratory of Modern Astronomy and Astrophysics
(Ministry of Education), Nanjing University, Nanjing 210093, China}

\altaffiltext{3}{Department of Computational and Data Sciences,
George Mason University, 4400 University Drive, MSN 6A2, Fairfax, VA
22030}

\begin{abstract}

In this paper, we analyze the full evolution, from a few days prior
to the eruption to the initiation, and the final acceleration and
propagation, of the CME that occurred on 2008 April 26 using the
unprecedented high cadence and multi-wavelength observations by
STEREO. There existed frequent filament activities and EUV jets
prior to the CME eruption for a few days. These activities were
probably caused by the magnetic reconnection in the lower atmosphere
driven by photospheric convergence motions, which were evident in
the sequence of magnetogram images from MDI (Michelson Doppler
Imager) onboard SOHO. The slow low-layer magnetic reconnection may
be responsible for the storage of magnetic free energy in the corona
and the formation of a sigmoidal core field or a flux rope leading
to the eventual eruption. The occurrence of EUV brightenings in the
sigmoidal core field prior to the rise of the flux rope implies that
the eruption was triggered by the inner tether-cutting reconnection,
but not the external breakout reconnection. During the period of
impulsive acceleration, the time profile of the CME acceleration in
the inner corona is found to be consistent with the time profile of
the reconnection electric field inferred from the footpoint
separation and the RHESSI 15-25 keV HXR flux curve of the associated
flare. The full evolution of this CME can be described in four
distinct phases: the build-up phase, initiation phase, main
acceleration phase, and propagation phase. The physical properties
and the transition between these phases are discussed, in an attempt
to provide a global picture of CME dynamic evolution.

\end{abstract}


\section{Introduction}

Coronal mass ejections (CMEs) are large-scale activities releasing a
vast amount of plasma and solar energetic particles (SEPs) into the
outer space \citep{gos93,webb94}. These plasma and SEPs can
propagate into the magnetosphere near the Earth and severely affect
space-based modern technological systems, especially during the
solar maximum \citep{smart89}. Solar physicists have been pursuing
what happens prior to the CME initiation and how CMEs are initiated.
Various observational signatures, including magnetic cancellation,
magnetic flux emergence, sigmoids, and filament activities are
regarded as the significant precursors of CME eruptions
\citep{martin98,canfield00,wang06,gibson2006}. The common nature of
these signatures are magnetic free energy build-up in the corona. As
a consequence of the energy build-up, the coronal magnetic fields
may explosively erupt once a trigger leads to the loss of
equilibrium \citep{forbes06}. However, there is no consensus so far
on the exact trigger mechanism. The MHD instability model suggests
that the eruption of CMEs that have a flux rope morphology is
probably caused by the kink and/or torus instability when the
winding of the field lines exceeds a critical value
\citep{sturrock01,linker01,fan04,rust05,totok05,gibson06}. In the
tether-cutting model, the magnetic reconnection that occurs close to
the polarity inversion line plays the role of weakening the
constraining tension force of the overlying field, and results in
the rise of the sigmoid-shaped core field and subsequently the
runaway eruption
\citep{moore80,moore01,sturrock89,liu07,sterling07}. The same
tension reduction mechanism holds for the flux emergence model
suggested by Chen et al. (2000) in which the magnetic reconnection
occurs between the emerging field and the background field. In the
breakout model proposed by Antiochos et al. (1999), the overlying
magnetic field constraining the sheared core field is removed
through external magnetic reconnection, which leads to the CME
eruption. Other authors have proposed that the injection of poloidal
magnetic flux (of sub-photospheric origin) in the flux rope can
cause a CME to take off \citep{chenj00,chenj03,krall01}. More
details about CME initiation mechanisms can be found in the reviews
(e.g., Forbes 2000, 2006; Gopalswamy 2003; Chen 2008; Schrijver
2009).

One key aspect of understanding CME eruption is of understanding the
relationship between CMEs and flares, which itself has been a
long-standing elusive issue for decades
\citep{kahler92,gos93,hund99}. Zhang et al. (2001, 2004) proposed
three phases of CME kinematic evolution: the initiation phase,
impulsive acceleration phase, and propagation phase, which are
tightly associated with the three phases of the associated flare:
the pre-flare phase, flare rise phase, and flare decay phase,
respectively (see also, Burkepile et al. 2004; Vr\v{s}nak et al.
2005b). The temporal correlation between CME acceleration and flare
HXR flux was studied by Qiu et al. (2004) and Temmer et al. (2008).
In the standard CME-flare model, the flare ribbons separate in the
chromosphere during the CME impulsive acceleration phase because of
continuous magnetic field reconnection. The reconnection rate can be
calculated in terms of flare ribbon separation speed and the
line-of-sight component of magnetic fields
\citep{forbes84,poletto86,forbes00,qiu02}. Qiu et al. (2004)
compared the reconnection rate with the acceleration of the
filament/CME and found a similarity between them. It was also found
that the total reconnection flux is proportional to the maximum
speed of CMEs (e.g., Qiu et al. 2005). In addition, Liu et al.
(2009) found that the spectral index of X-ray emission of flares is
strongly anti-correlated with the reconnection electric field. All
of these suggest that CMEs and the associated flares, during the
impulsive energy-release phase in particular, are driven by the same
physical process in the lower corona, presumably via magnetic
reconnection \citep{lin00,priest02,vr04b,zhang06,mari07,temmer08}.

As a matter of fact, most previous studies concerning CMEs address
only certain specific phases of CME evolution, while very few are
for the full evolution cycle from the build-up phase (tens of hours
prior to the CME initiation), throughout the initiation phase and
acceleration phase, and to the propagation phase. Therefore, it is
useful to make a complete observation to investigate the full CME
evolution. It is also of particular interest to study the variation
of magnetic topology involved in different evolution phases, which
shall shed light on possible initiation mechanisms of CMEs. The
unique data of high cadence and full coverage acquired by SECCHI
(Sun Earth Connection Coronal and Heliospheric Investigation; Howard
et al. 2008) instruments onboard STEREO (Solar Terrestrial Relations
Observatory; Kaiser et al. 2008) spacecraft provide us the
opportunity to make such a study. In this paper, we investigate the
full evolution of the CME on 2008 April 26 which was well observed
by STEREO. In $\S2$, we describe the instruments and the data. Our
analysis and results are shown in $\S3$ and $\S4$. In $\S5$, a
schematic model is proposed to explain the full evolution of the
CME, followed by discussions and conclusions in $\S6$.

\section{Instruments and Observations}

The STEREO spacecraft were designed to monitor solar activities from
two different perspectives in space, which for the first time
provide the stereoscopic measurement for understanding solar
eruptions. STEREO A moves ahead of the earth in its orbit and STEREO
B trails behind. The separation between the two spacecraft has been
continuously increasing. In particular, the separation angle was
49$^{\circ}$.5 on 2008 April 26 so that the CME studied in this
paper were observed from two well-separated perspectives.

The SECCHI instrument suite on board STEREO is composed of five
telescopes. Most of the data analyzed in this paper are from three
of them: Extreme UltraViolet Imager (EUVI), Inner Coronagraph
(COR1), and Outer Coronagraph (COR2). EUVI observes solar
chromosphere and the lower corona at four passbands: 171 {\AA}, 195
{\AA}, 284 {\AA}, and 304 {\AA}, with a cadence higher than EIT
(Extreme-ultraviolet Imaging Telescope; Delaboudini\`{e}re et al.
1995) onboard SOHO, especially at 171 {\AA} passband. COR1 and COR2
are externally occulted white-light coronagraphs with fields of view
(FOVs) of 1.4--4.0$R_\odot$ and 2.5--15.6$R_\odot$, respectively.
Both of them have a cadence higher than LASCO (Large Angle and
Spectrometric Coronagraph; Brueckner et al. 1995) on board SOHO.
Therefore, the SECCHI instruments can well observe CMEs from its
birth place on the solar surface to its ultimate propagation in the
outer corona.

In addition, GOES X-ray data reveal the temporal profile of the soft
X-ray emission at 1--8 {\AA} for solar-flares, which are often
associated with CMEs. The RHESSI (Reuven Ramaty High Energy Solar
Spectroscopic Imager; Lin et al. 2002) spacecraft provides the HXR
light curve of the flares. The location of flares can be found in
Solar Geophysics Data
Reports\footnote{http://www.ngdc.noaa.gov/stp/SOLAR}. The MDI
(Michelson Doppler Imager; Scherrer et al. 1995) images provides the
longitudinal magnetic field at the surface of the Sun.

Up to date, more than three hundred CMEs have been observed by
STEREO since the launch in 2006
December\footnote{http://cor1.gsfc.nasa.gov/}. The CME that occurred
on 2008 April 26 was associated with a GOES B3.8 class flare. It
appeared near the solar limb as seen from the perspective of STEREO
A while near the disk center as a halo CME as seen from the
perspective of STEREO B. From inspecting the running difference
images, the leading edge (LE) of the CME that appeared in the FOVs
of COR1 and COR2 was evidently sharp, thus in favor of the height
measurement. From the height measurement, we can further infer the
CME's kinematical evolution. Recently, Thernisien et al. (2009)
obtained its 3D velocity and average acceleration by a forward
modeling method using only the SECCHI/COR2 observations. Here, we
study this event using multi-wavelength data with a focus on its
full evolution, including the magnetic topology in different phases.

\section{Precursors, Initiation, and Eruption of the CME}

\subsection{Magnetic Cancellation in the Active Region}
Inspecting the surface source region in magnetogram images, we found
a continuous flux cancellation for several days prior to the
eruption. We plotted in Figure 1 the line-of-sight (LOS) magnetic
field of the active region in which the CME was originated. The
three magnetograms in Figure 1 were taken in three consecutive days
prior to the CME eruption; they had been rotated to the same time in
order to have a better comparative view. The cancellation of
magnetic flux mainly occurred near the polarity inversion line
(PIL). We showed in Figure 2 the changes of the magnetic fluxes in
both the whole region (left panel) and the central region (indicated
by the white rectangle in Figure 1). Both the positive flux and the
negative flux decreased slowly, as well as the unsigned magnetic
flux in the whole region. For the central region near the PIL, the
positive magnetic flux decreased sharply from about 5.8 $\times$
10$^{20}$ to 2.2 $\times$ 10$^{20}$ Mx during the three days before
the eruption. Note that the increasing of negative magnetic flux for
about one day prior to the onset of the CME/flare (denoted by the
vertical solid line in Figure 2), was due to the motion of the
negative patch toward the northeast; the change was caused by flux
transportation instead of flux emergence. Therefore, the slow
converging motion between the two opposite polarities, driven
probably by photospheric flows, resulted in the continuous magnetic
cancellation. We believe that this photospheric flux cancellation
process lead to subsequent filament activities and ultimately the
CME we will be discussing. The magnetic cancellation has been
considered as one of primary magnetic signatures leading to major
solar activities (e.g., Wang et al. 2006).

\subsection{Disappearance of the Associated Filaments}
We carefully examined what occurred in the corona of the source
active region within two days prior to the CME eruption. As shown in
the left and middle panels of Figure 3, there appeared the existence
of a filament over the active region. It was initially visible for
several hours on April 24, during which several jets occurred,
indicated by the white arrow in the left panel. The filament was
also visible on April 25 during which some mass flew down slowly
along the field lines. Eventually, the filament rose impulsively on
April 26 (indicated by the white arrow in the right panel of Figure
3). Note that such active behaviors of filaments before their final
eruption has been well observed before (see the review by Pick et
al. 2006). Nevertheless, the final eruption of filaments is closely
associated with that of CMEs.

\subsection{Sigmoid Configuration and EUV Brightening}
The CME of interest also had the sigmoid signature prior to the
eruption. Figure 4 showed some sampling EUV images of the active
region prior to the CME eruption; the sigmoid was particularly
obvious in the 284 {\AA} (lower-middle panel) and 195 {\AA}
(lower-right) passband images. The 171 {\AA} passband images, on the
other hand, did not show the sigmoid well, but revealed better the
morphology of the overlying magnetic loop arcade (upper panels). The
sigmoid seemed consisting of two co-existing J-shaped bundles of
low-lying loops forming a reversed-S shape in projection; the two
ends were in opposite sides of the PIL and anti-parallel to each
other. Similar sigmoid structures comprising of many individual
loops have been identified by McKenzie et al. (2008). Such
structures are preferentially observed in eruption regions and thus
have been regarded as a precursor of CMEs
\citep{canfield00,gibson2006}. Note that, the 195 {\AA} image at
9:06 UT on April 25 (lower-left panel) showed the coronal
configuration prior to the formation of the coronal sigmoid.

Another noticeable feature was the EUV brightening in the sigmoid
core field, most clearly seen in the 171 {\AA} passband images, as
denoted by the two small squares in Figure 4 (upper panels). The
brightening can also be seen in the 284 and 195 {\AA} passband
images. Such brightening, first appearing at $\sim$12:40 UT, implies
that magnetic reconnection occurred at this site about one hour
before the eruption. We also note that some twisted field lines in
the sigmoid started to rise slowly from $\sim$13:36 UT. These
observations help understand the triggering or initiation mechanism
of the CME. They seem to favor the tether cutting model proposed by
Moore et al (2001) and further elaborated by Liu (2007), but not the
breakout model proposed by Antiochos et al. (1999). First, the
active region on the photosphere appeared as a simple bipolar
magnetic field, and the corona magnetic field could be characterized
by a sigmoidal core field constraining by an overlying bipolar
arcade field. Secondly, the pre-eruption EUV brightenings only
occurred within the core field; we do not find any remote
brightenings surrounding the active region as expected from the
breakout model \citep{moore06}. The breakout model usually requires
a quadrupole magnetic configuration and an initiation magnetic
reconnection at the null point above the central core field.
Therefore, the observational features of this event are well
consistent with the tether-cutting model, in which the magnetic
reconnection occurs within the low-lying core field lines
\citep{moore01}. We believe that the initial reconnection, as
indicated by the EUV brightenings, caused the slow rise of the
sigmoidal magnetic structure.

\subsection{Eruption of the CME}

From inspecting the EUVI movies, we found that the overlying field
lines were relatively stable for days before the eruption at about
13:44 UT, at which the whole system became unstable and erupted
impulsively. The CME first appeared in the FOV of the COR1 A image
at 14:15 UT at a height of 1.91$R_\odot$ from the disk center. From
the COR1 A movie, one can clearly see that a sharp semicircular
front expanded outward. Whereas, the CME first appeared in the FOV
of the COR1 B image at 14:25 UT as a halo shape at a height of
1.72$R_\odot$. At 14:55 UT, another inner halo structure appeared.
The two halos may correspond to the CME disturbance front and the
expanding flux rope, respectively (see also, Wood \& Howard 2009).
Selected snapshots of the CME were shown in Figure 5. Note that
during the CME eruption, a streamer disturbance was triggered at the
southern side as indicated by the black arrow in Figure 5.

\section{Kinematics of the CME}

We can well track the LE of the CME using the running difference
images at 171 {\AA} and in white light images, as shown in Figure 5.
The white arrows in the figure pointed to the position angle of the
measurement, at which we measured the height-time variation of the
CME. The height-time measurement was then used to derive the
velocity profile through the piece-wise numerical derivative method,
i.e., the Lagrangian interpolation of three neighboring points
\citep{zhang01,zhang04}. From the velocity profile, the CME
acceleration can be further derived through a similar method but
having a larger uncertainty. We thus obtained the full kinematic
evolution of the event from the solar surface continuously to the
outer corona as indicated in the middle panels of Figure 7. Note
that the uncertainty in the height measurements was estimated to be
0.026, 0.12, and 0.24 $R_\odot$ for EUVI, COR1, and COR2,
respectively. This is the main factor causing the uncertainty in the
calculation of speed and acceleration of a CME \citep{zhang04}.

The reconnection electric field can be calculated using the
separation speed of the H$\alpha$ ribbons, serving as a proxy of the
magnetic reconnection rate. Here, we revisited this issue using EUVI
observations. For the event here, the flare ribbons were well shaped
and thus can be easily traced. We chose five directions, marked by
the white lines in Figure 6, to measure the separation speed of the
flare ribbons. The final separation speed was obtained by average
the speeds along the five lines. The magnetic field was taken from
the MDI magnetogram image just prior to the CME eruption. Since the
event occurred near the disk center as seen from SOHO, the observed
LOS magnetic field should be close to the radial component of the
fields needed for the calculation. The reconnection electric field
can then be inferred as $E_{rec} = \overline{V} B_{n}$, which was
plotted as the green lines in the bottom panels of Figure 7.

The relationship between the acceleration phase of the CME and the
impulsive phase of the associated flare had been investigated in
many papers (e.g., Zhang et al. 2001, 2004, Temmer et al. 2008). Qiu
et al. (2004) and Jing et al. (2005) compared in detail the
acceleration of the CME, the reconnection rate, and the HXR emission
of the associated flare. In the two studies, they used the filament
acceleration as the proxy of CME acceleration because of the lack of
CME observations in the inner corona. However, thanks to the STEREO
observations, we were able to study the event on 2008 April 26 with
better continuity in space and time. We derived the CME's
acceleration from two well-separated viewing angles and tracked the
CME continuously from the solar surface to the outer corona. We then
compared the CME' acceleration with the reconnection rate and the
HXR flux of the associated flare, as shown in the bottom panels of
Figure 7. The acceleration of the CME peaked at 13:51 UT, while the
reconnection electric field and the RHESSI 15--25 keV HXR flux
peaked at 13:54 UT. This time difference was relatively small and
was within the time resolution of the CME acceleration curve. In
general, the profile of CME acceleration was coincident with the
profile of the reconnection electric field and the HXR flux curve.

\section{A Four-Phase CME Evolution Model}

In this section, we attempt to use a schematic model to explain the
full evolution of the CME from the early development to the ultimate
eruption (Figure 8). We piece together many components proposed by
other researchers which we believe were relevant to CME evolution
and put them into a coherent scenario. For the sake of clarity of
discussion, we suggest that the full evolution of the CME should be
divided into four phases: (1) the build-up phase, (2) the initiation
phase, (3) the main acceleration phase, and (4) the propagation
phase.

The build-up phase was the phase of preparation that lasted for
days. As discussed earlier, it was characterized by many pre-cursor
signatures: flux cancellation, filament activity, sigmoid and EUV
brightening, even though these signatures were neither necessary nor
sufficient for an eruption. Formation and evolution of filaments had
been extensively studied for many years. Martens et al. (2001)
proposed a head-to-tail model to explain the formation of filaments
through flux convergence and cancellation. Subsequently, Welsch et
al. (2005) simulated the filament formation using two flux systems
driven by the convergence of opposite polarities along PILs. Chae et
al. (2001) proposed that slow magnetic reconnection last all the
time in the chromosphere driven by converging motions. The
continuous reconnection can result in both the overlying field lines
straddling the neutral line and the low-lying core field lines
\citep{chae01,welsch05}.

Further, some EUV jets and small eruptions took place at the site of
magnetic cancellation. As the positive and negative fluxes moved
close to each other near the PIL, the anti-parallel inner ends of
the two bundles of the loops reconnected slowly and continuously in
the lower atmosphere (i.e., the chromosphere). That resulted in the
formation of the overlying M-shaped field lines, almost
perpendicular to the PIL. At the same time, it also produced the
low-lying field lines, near parallel with the PIL. The active
filaments were condensed at the dip of the M-shaped field lines, as
indicated in the upper right panel of Figure 8. As the filament mass
flew down along the M-shaped field lines, more field lines rose and
served as the overlying loops. These loops were heated slowly and
remained to be invisible at 171 {\AA} until about tens of hours
prior to the CME eruption. Note that, as long as some open field
lines exist at the reconnection site, part of the filament mass may
erupt as EUV jets. However, although part of the filament mass flew
down slowly or erupted, the rest of the filament appeared to be
quite stable in the dip of the field lines all the time prior to the
CME eruption (Figure 8c). With time going on, the lower field lines
in the eastern part and the pre-existing field lines in the western
part, being both J-shaped, moved closer to each other, driven by the
continuous convergence motion along the PIL and formed a reversed-S
sigmoid structure in the projection plane. Then, the ends of the two
bundles of the J-shaped loops, on the opposite sides of the PIL,
reconnected as tether-cutting and formed the little twisted field
lines, while the energy released through the reconnection heated the
plasma in the middle part of the reversed-S sigmoid configuration
and thus producing the observed EUV brightenings. The shortest field
lines submerged into the sub-photosphere after the slow
reconnection, which was manifested by the magnetic cancellation in
the photosphere (Figure 8b--d). Recently, Tripathi et al. (2009) and
Green \& Kliem (2009) also reported such sigmoid structure coming
into existence after a pair of J-shaped arcs reconnecting through
flux cancellation in the photosphere. It should be noted that these
sigmoid structures may provide observational evidence of the flux
rope existence prior to the CME eruption, which was useful for
distinguishing different CME-flare models.

The initiation phase occurred when the upward force within the
sigmoid was able to overcome the tension force of the overlying
field lines. As more and more J-shaped loops reconnected by
tether-cutting, the twisted field lines in the reversed-S sigmoid
configuration beneath the overlying loops moved up due to an
increased upward magnetic hoop force and a decreased downward
magnetic stress \citep{moore01,liu07,sterling07}. The rising twisted
field lines pushed upward the overlying loops. When the overlying
loops were stretched to a certain extent due to the tether cutting
reconnection, a current sheet between the legs of the distended
overlying field line was formed under the loops so that a fast
runaway reconnection was subsequently initiated, leading to the main
energy release phase and the impulsive acceleration of the CME; this
is the standard model of eruptive flares \citep{Hirayama74}. Another
possibility leading to the main eruption is the triggering of MHD
instability of the flux rope formed from the tether cutting
reconnection, through kink and/or torus instability
\citep{totok05,Kliem06}.

The subsequent main acceleration phase is believed to be caused by
the runaway magnetic reconnection, coupled with the explosive
poloidal flux injection into the rising flux rope. The reconnection
rapidly injected a large amount of poloidal flux into the twisted
field lines, thus supplied a stronger upward driving force so as to
impulsively accelerate the CME flux rope. On the other hand, the CME
eruption led to a decrease of the magnetic pressure below the flux
rope, which caused a faster inflow toward the current sheet and
enhanced the runaway reconnection. This positive feedback process
effectively released the magnetic free energy stored in the lower
corona, which converted into the kinetic energy of the CME and also
produced the enhanced soft X-ray and hard X-ray emissions
\citep{li93}. Moreover, the CME eruption led to a depletion of mass
in the lower atmosphere near the active region and formed the
coronal dimming \citep{thompson98}. As the magnetic reconnection
progressed, the reconnection site rose gradually. The upward moving
reconnection site induced the flare ribbons to separate horizontally
at the base of the corona, as evidently seen in the EUV and
H$\alpha$ channels. Beneath the reconnection site, the newly
reconnected magnetic loops were filled by the plasma evaporated from
the chromosphere and the sigmoid magnetic structure changed to
post-flare loop arcades (see also, Liu et al. 2007), as shown in
Figure 8f. Note that the magnetic configuration before the eruption
(Figure 8c) had been unambiguously confirmed by the extrapolated
coronal magnetic field obtained from a linear force-free field
(LFFF) model using the MDI magnetogram as the input.

After the main phase that lasted about 10 minutes, the runaway
reconnection came to a stop. The CME now entered into the simple
propagation phase: the CME was propagating in nearly a constant
speed or with a small residual acceleration in the outer corona.

\section{Discussions and Conclusions}

The STEREO observations provide an unprecedented opportunity to
investigate solar eruptions. In this paper, we presented
multi-wavelength observations of the flare-associated CME that
occurred on 2008 April 26. We had studied its evolution for a long
period and discussed the full evolution in a four-phase scenario:
the build-up phase, initiation phase, main acceleration phase, and
propagation phase. During the build-up phase, the active filaments,
instantaneous EUV jets, and a reversed-S sigmoid structure were
observed. All the features were physically related to the persistent
slow magnetic reconnection in the solar lower atmosphere, which was
manifested as photospheric magnetic cancellation. Before the
eruption, there was a long period of reconnection occurring in the
lower layers resulting in the transferring and accumulation of
magnetic free energy, as well as the formation of a magnetic
structure favorable for eruption, i.e., the sigmoid structure in
this event. Different from the process of flux cancellation, the
emerging of magnetic flux may also play an important role in
transferring magnetic free energy from the sub-photosphere into the
corona \citep{tian08,Archontis09}. MacNeice et al. (2004) showed
that, as the magnetic field shear increases, the magnetic free
energy is continuously accumulated. They also proposed that such a
quasi-static energy accumulation phase is necessary for any fast CME
eruption. The magnetic field shear can be caused by the convergence
motion of opposite magnetic fluxes \citep{titov08}. Using a
nonlinear force-free field extrapolation, it was recently found that
the accumulated magnetic free energy increases with time prior to
the eruption \citep{thalmann08,guo08,jing09}. Therefore, the
build-up phase accumulates the sufficient magnetic free energy for
the eventual initiation and the final eruption.

In general, the initiation phase of a CME eruption is characterized
by a slow rise of the CME flux rope. In the present event, the EUV
emission started to brighten in the core part of the reversed-S
sigmoid configuration, which implied that slow magnetic reconnection
was taking place there. As the field lines in the reversed-S sigmoid
configuration continually reconnected, the CME flux rope rose slowly
for about 20 minutes. This inner core magnetic reconnection prior to
the eruption, combined with the facts of the bipolar magnetic
structure in the active region and the absence of remote
brightenings, seems to rule out the breakout model as the trigging
mechanism of this event. Instead, we think that this eruption is
well consistent with the tether-cutting initiation model. We also
investigated the kinematics of this CME and found that its
acceleration was well correlated with the HXR flux of the associated
flare and the magnetic reconnection rate (see also, Zhang et al.
2004; Qiu et al. 2004; Temmer et al. 2008). It suggests that the
main acceleration phase of the CME in the inner corona is likely
caused by the fast runaway magnetic reconnection. Later on, the CME
propagated with almost a constant velocity in the outer corona (see
also, Zhang et al. 2001, Gallagher et al. 2003). However, for CMEs
associated with a long decay flare, even though the fast magnetic
reconnection ceases, a positive post-impulsive-phase acceleration
may continue to exist after the impulsive acceleration phase
\citep{cheng09}. In general, these observational results are
consistent with the standard CME-flare model.

We think that the schematic model that comprises of the four phases
proposed in this paper can be applied to most CME events. However,
owing to different physical circumstances under which CMEs occur,
individual events may have their own characteristics. In particular,
there may be various manifestations for the build-up phase and the
initiation phase, as mentioned above. We look forward to more
observations in the coming years to study a variety of CME events,
in order to fully understand the full evolution cycle of CMEs,
including energy build-up, initiation, impulsive acceleration and
subsequent propagation in the interplanetary space.

\acknowledgments

We thank to the referee for the valuable suggestions and comments
that helped to improve the paper significantly. We are grateful to
P. F. Chen, Z. J. Ning, and C. Liu for valuable discussions, and to
M. Jin for help on data analysis. We thank the STEREO/SECCHI data
provided by a consortium of NRL (US), LMSAL (US), NASA/GSFC (US),
RAL (UK), UBHAM (UK), MPS (Germany), CSL (Belgium), IOTA (France),
and IAS (France). This work was supported by NSFC under grants
10673004, 10828306, and 10933003 and NKBRSF under grant
2006CB806302. SOHO is a project of international cooperation between
ESA and NASA.

\begin{figure}
\epsscale{1.00} \plotone{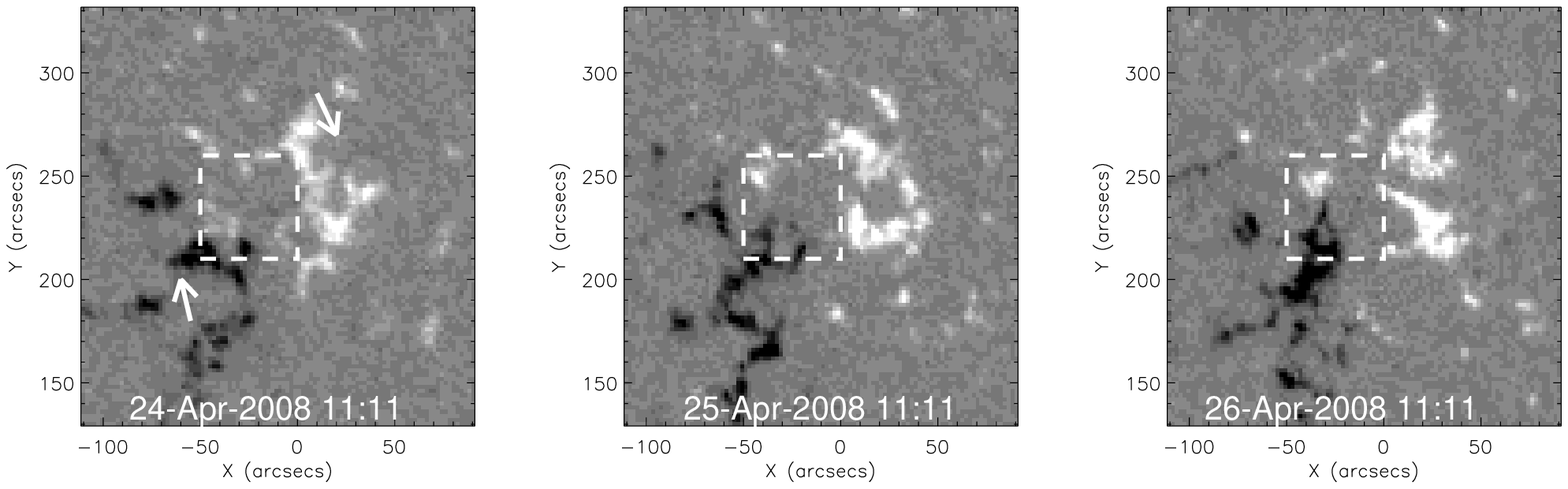} \caption{Line-of-sight
magnetograms with a field of view of 200\arcsec $\times$ 200\arcsec
taken before the CME eruption. The white square denotes the place of
the core field near the PIL, where the main flux cancellation
occurred. The two arrows show the motion direction of the two
opposite polarities. \label{fig1}}
\end{figure}

\begin{figure}
\epsscale{1.0} \plotone{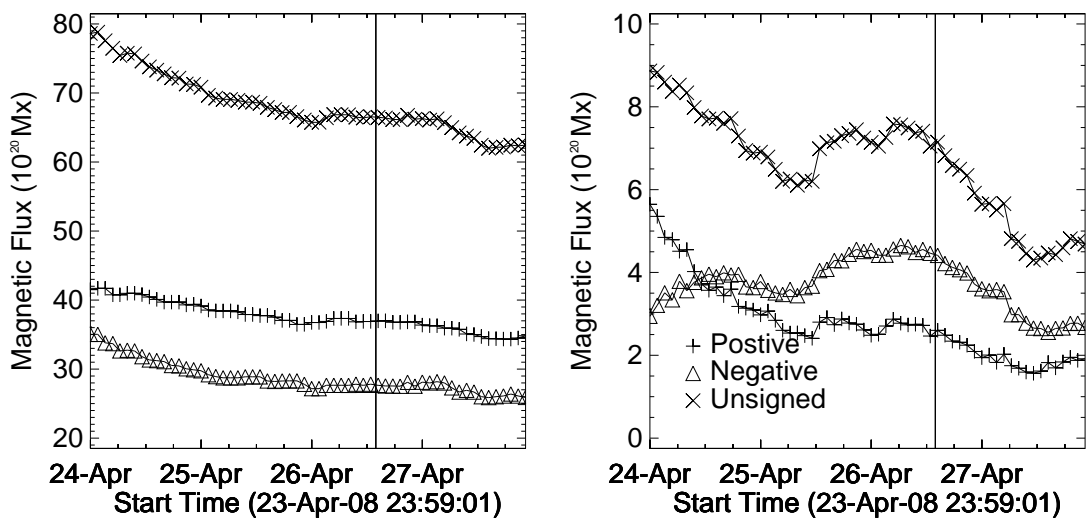} \caption{The magnetic flux
evolution for the whole FOV (left) and that for within the white
square (right) in Figure 1. Note that the increase of the negative
flux in the right panel is mainly caused by the transportation of
negative patch toward northeast. The vertical line denotes the onset
of the CME. \label{fig2}}
\end{figure}

\begin{figure}
\epsscale{1.0} \plotone{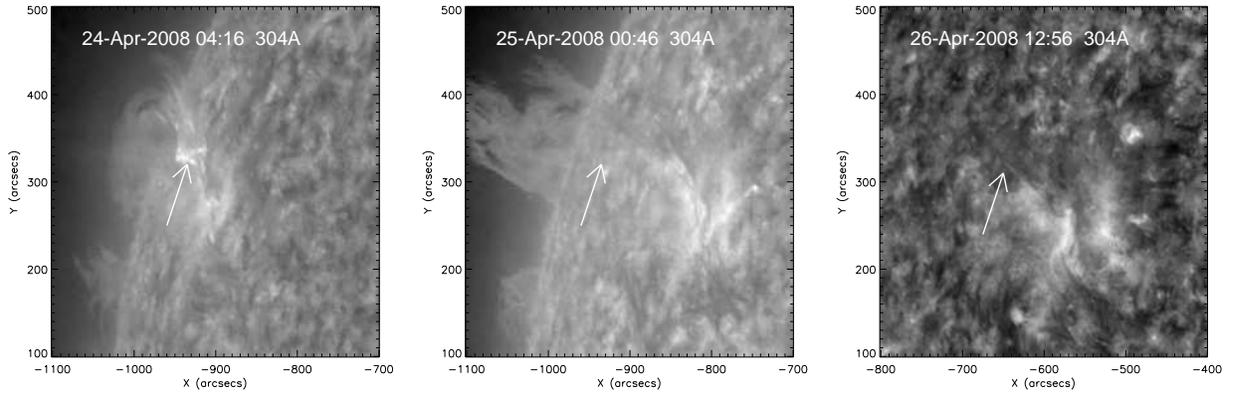} \caption{EUVI-A 304 {\AA}
images showing active filaments and jets before the CME eruption.
\label{fig3}}
\end{figure}

\begin{figure}
\epsscale{1.00} \plotone{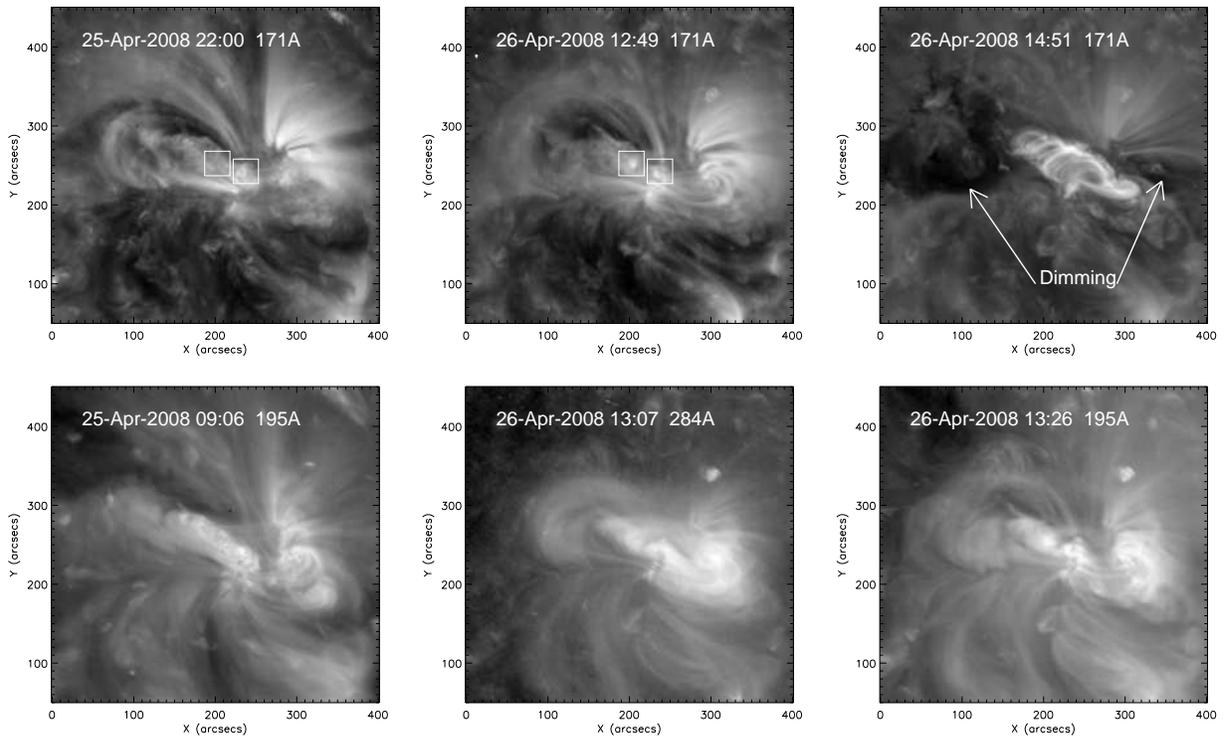} \caption{EUVI-B images in three
different passbands showing the overlying loops and the sigmoid
configuration before the CME eruption. The small white squares show
the brightening of the core field. The arrows denote the dimming
regions.\label{fig4}}
\end{figure}

\begin{figure}
\epsscale{1.00} \plotone{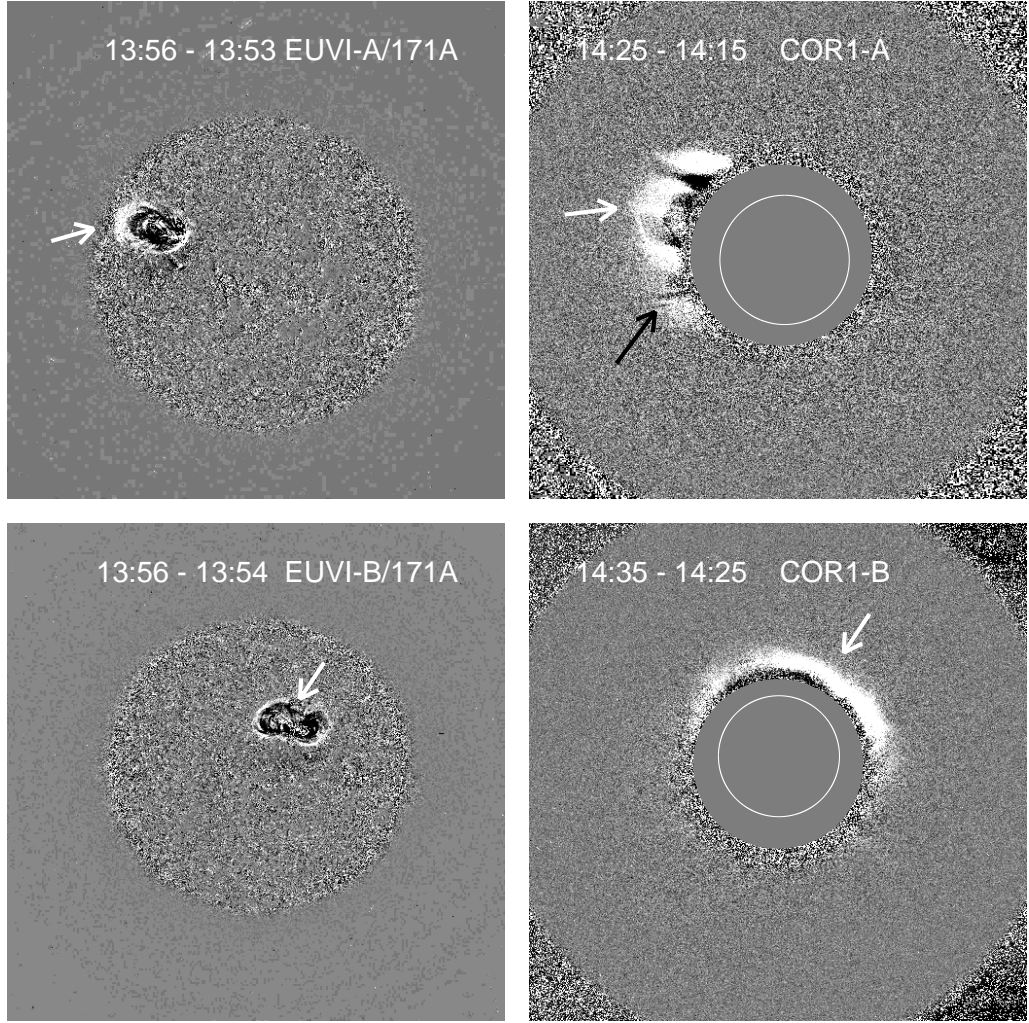} \caption{STEREO A (top panels) and
B (bottom panels) running difference images. The white arrows
indicate the CME feature at which we measure the height-time
evolution. The black arrow in top-right panel indicates a disturbed
streamer. \label{fig5}}
\end{figure}

\begin{figure}
\epsscale{0.5} \plotone{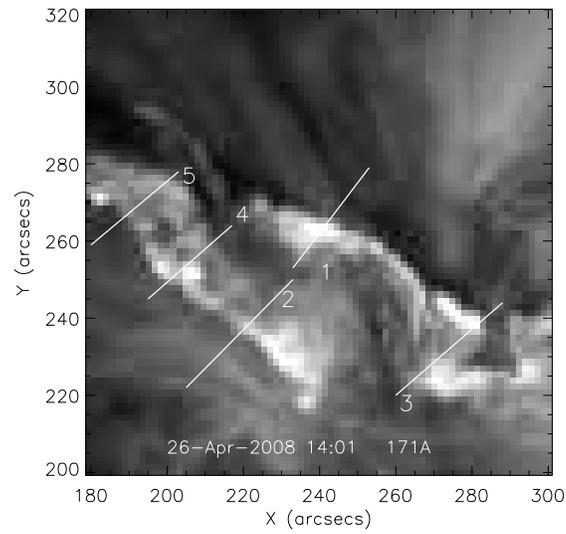} \caption{An image in 171 {\AA}
passband showing flare ribbons. The five lines refer to the
directions along which we measure the separation speed of the
ribbons. \label{fig6}}
\end{figure}

\begin{figure}
\epsscale{1.0} \plotone{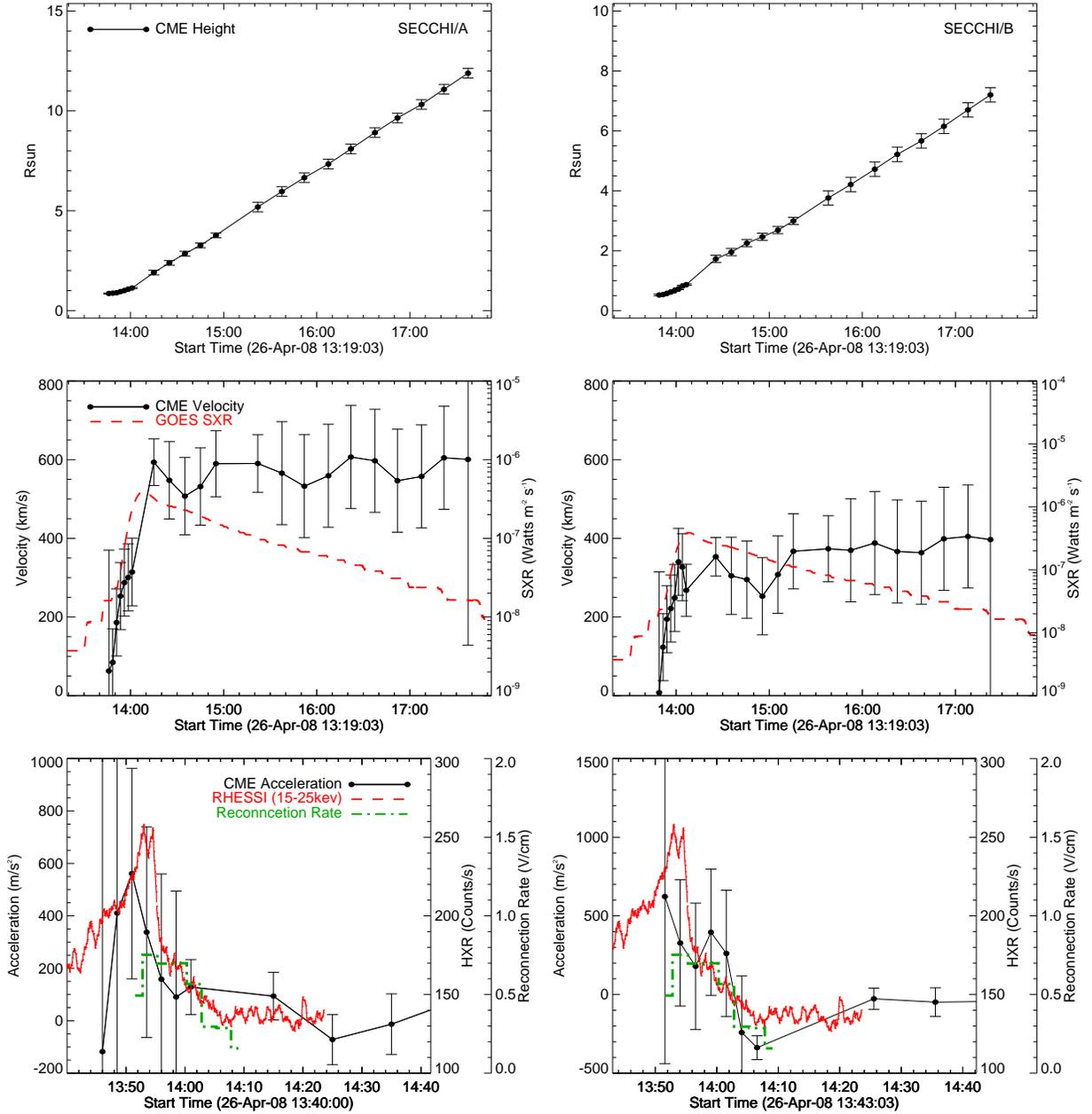} \caption{The temporal
evolution of the height (top), speed (middle), and acceleration
(bottom) of the CME obtained from STEREO A (left) and B (right). The
GOES SXR and RHESSI HXR flux profiles of the associated flare are
shown in the middle panels and the bottom panels, respectively. Also
shown in the bottom panels is the reconnection electric field.
\label{fig7}}
\end{figure}

\begin{figure}
\epsscale{1.00} \plotone{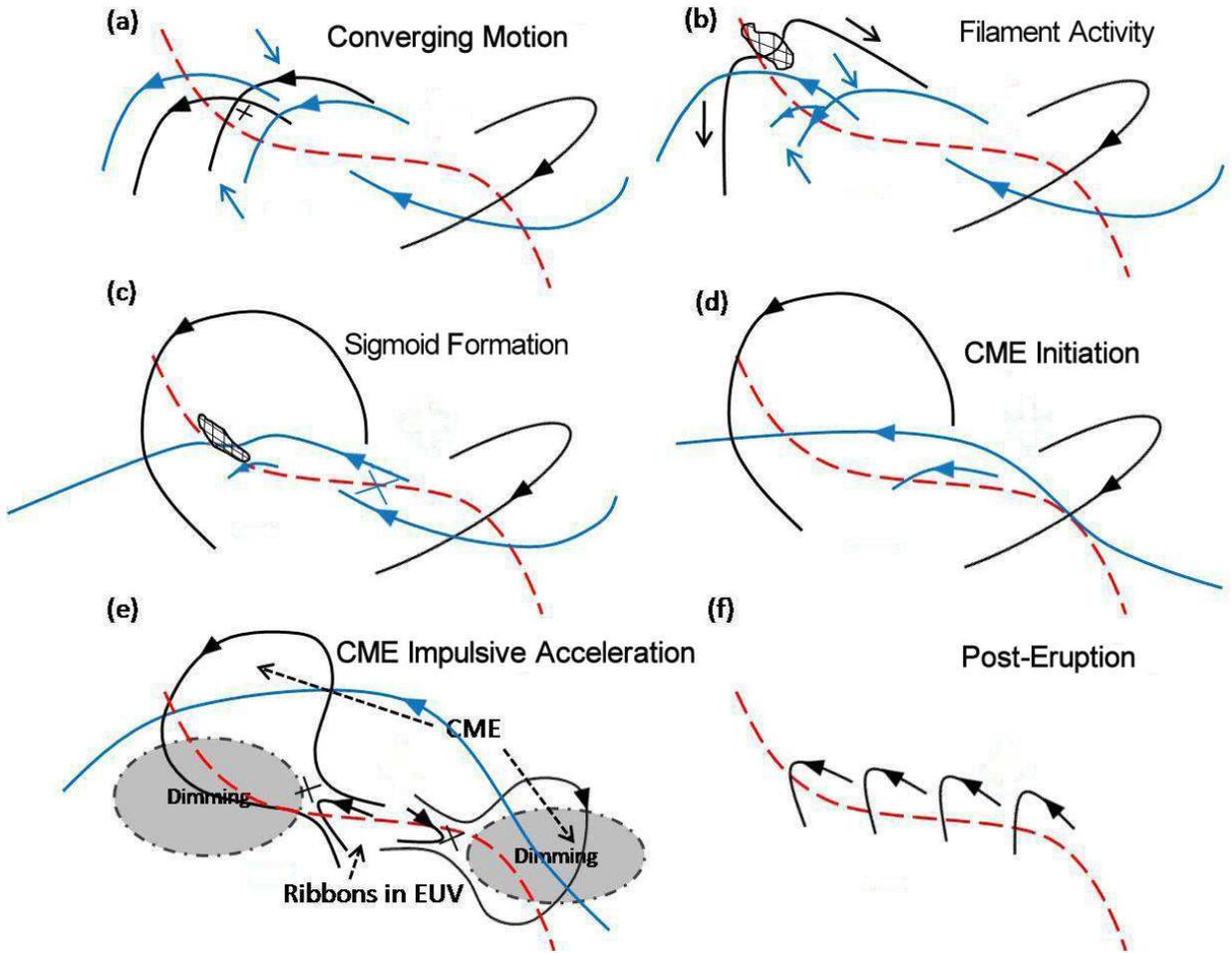} \caption{A schematic drawing
of selected magnetic field lines illustrating the evolution of the
event. Black lines refer to the overlying magnetic field and blue
lines to the the axis of the flux rope, both of which change
obviously in different phases. The red dashed line is the polarity
inversion line. The cross denotes the reconnection site.
\label{fig8}}
\end{figure}

\end{document}